\documentclass[12pt,preprint]{aastex}
\usepackage{graphicx,epsfig,epsf,amssymb}
\def \<{\langle}
\def \>{\rangle}
\def \tt{$\sqrt{\<\hat{t}^2\>}$}

\newcommand{\Hb}{H$\beta$ }
\newcommand{\Ha}{H$\alpha$ }
\newcommand{\Hd}{H$\delta$ }
\newcommand{\Hg}{H$\gamma$ }
\newcommand{\mc}{\multicolumn}

\begin{document}

\title{Five E+A (post-starburst) galaxies as Brightest Cluster Galaxies}

\author{Feng-Shan Liu\altaffilmark{1,2}, 
Zhong-Lue Wen\altaffilmark{2}, 
Jin-Lin Han\altaffilmark{2}
and Xian-Min Meng\altaffilmark{2}}

 \altaffiltext{1}{
College of Physical Science and Technology,
Shenyang Normal University, Shenyang, 110034, China; e-mail: lfs@nao.cas.cn}
 \altaffiltext{2}{
National Astronomical Observatories, Chinese Academy of Sciences,
A20 Datun Road, Beijing, 100012, China
}

\begin{abstract}
Brightest Cluster Galaxies (BCGs) are mostly elliptical galaxies
and very rarely have prominent star formation. We found that five
out of 8,812 BCGs are E+A (i.e. post-starburst) galaxies, having
the \Hd~absorption line with an equivalent width $>2.5\AA$ and no
distinct emission lines in [O II] and \Ha. The E+A features we
identified from the BCGs for the first time are not as significant
as those in general galaxies, indicating that historically the
star formation were not very violent.
\end{abstract}
\keywords{galaxies: active -- galaxies: starburst -- galaxies: formation }

E+A galaxies have strong Balmer absorption lines in their spectra
but no distinct emission lines of [O II] at $\lambda$3727 and \Ha.
They look like a superposition of spectra of elliptical galaxies
and A-type stars (E+A). The strong Balmer absorption lines
indicate that these galaxies experienced a powerful star formation
within 1 Gyr, but no star formation is going on since
non-detection of [O II] emission line (e.g., Couch \& Sharples 1987; Poggianti et al. 1999).  
Therefore, E+A galaxies have been interpreted as post-starburst galaxies, which
is an important stage of overall galaxy evolution in the Universe.

E+A galaxies (also called as K+A galaxies) were first discovered
in distant clusters of galaxies (e.g., Dressler \& Gunn 1983). E+A galaxies now have been
found in various environments, very often in cluster regions of
redshifts $z>\sim0.4$ (e.g., Dressler et al. 1999; Tran et al. 2003). Low redshift E+A galaxies are found
predominantly in the field environment (e.g., Blake et al. 2004; Quintero et al. 2004). Most of E+A galaxies
are spiral galaxies but some are elliptical galaxies (e.g., Goto et al. 2003; Huang \& Gu 2009). Less
than 1\% of all galaxies in the present Universe are E+A galaxies (Goto et al. 2003). 
E+A galaxies may provide hints for evolution of early-type galaxies.

The Brightest Cluster Galaxies (BCGs) are the most luminous and
massive galaxies in the center of galaxy clusters. They are mostly
elliptical galaxies and dominated by old stars without prominent
ongoing star formation. However, they are different from ordinary
elliptical galaxies (non-BCGs) in the surface brightness profiles
and some basic scaling relations (see Liu et al. 2008 for a short review).
Because of the dominant role inside clusters and their unusual
properties, the formation and evolution of BCGs are very
intriguing.

Very rarely BCGs show signatures of ongoing star formation (e.g., Crawford et al. 1999; O'Dea et al. 2008). 
When the star formation in these BCGs is quenched, they
may evolve into the E+A stage.  There have been no reports
previously about E+A BCGs, probably due to their rarity.

We searched for the signatures of E+A galaxies from the BCGs of
SDSS-WHL clusters (Wen et al. 2009). Wen et al. (2009) have identified 39,668
clusters in the redshift range $0.05<z\lesssim0.6$, To find the
E+A spectral features, we checked the BCGs which have
spectroscopic observations and their spectra have been
parametrized by the MPA/JHU
team\footnote{http://www.mpa-garching.mpg.de/SDSS/DR7/}. Among
16,276 BCGs with spectroscopic spectra, we only looked at 12,186
BCGs in the redshift range of $0.05<z<0.4$ because the \Ha~line
otherwise goes outside the wavelength coverage of 3800-9200\AA~for
SDSS spectroscopy. We further discarded the spectra with a
median signal-to-noise (S/N) per pixel less than 5, and got only spectra of 8,812
BCGs.

Previously E+A galaxies are found solely based on the [O II] line
at $\lambda3727\AA$ and \Hd absorption line. Different criteria
were used in literature. The \Hd equivalent width (EW) is usually
set to be, e.g., $\rm EW(H_{\delta})>3$\AA~by Poggianti et al. (2004), 
$>$5\AA~by Goto (2007) and Falkenberg et al. (2009) and
$>$2.5\AA~by Huang \& Gu (2009). Note that the value for absorption
lines here has a positive sign, and that for emission lines will
set to have a negative sign. The [O II] EW is adopted to $\rm
EW([OII])$ $>-5$\AA~by Tran et al. (2003) and $>$-2.5\AA~by Goto (2007). 
Goto et al. (2003) showed such E+A galaxies without information
of \Ha line suffer from a large fraction ($\sim$52\%) of
contamination from \Ha emitting galaxies. Other Balmer absorption
lines (\Hg and \Hb) were also considered for high-redshift E+A
galaxies when the \Ha line is not available (e.g., Tran et al. 2003; Blake et al. 2004; Yang et al. 2004). 

We follow the criteria of Huang \& Gu (2009) to select E+A BCGs: the
$\rm EW(H_{\delta})>2.5\AA$, $\rm EW([OII])>-2.5\AA$ and $\rm
EW(H_{\alpha}) >-3\AA$. We also request the measured values of \Hd
EW and flux are better than $3\sigma$ (i.e. 3 times of
uncertainties). These criteria are reliable for selection for E+A
galaxies, because the inclusion of \Ha EW can exclude the
contamination from \Ha emitting galaxies efficiently. We also
exclude the BCGs at $0.35<z<0.37$ from our analysis because the
sky feature at 5577\AA~may affect the measurement of \Hd~line
(with the window of 4092 - 4111\AA~in MPA/JHU). Finally, among
8,812 BCGs, five are E+A BCGs with the \Hd~equivalent width
$>$2.5\AA~, [O II] equivalent width $>-2.5$\AA~and \Ha~equivalent
width $>-3$\AA.  Their spectra are shown in Figure 1, and basic
parameters are listed in Table 1. We checked the images of these
E+A BCGs and clusters carefully, and noticed that all E+A BCGs are
definitely the brightest galaxies in clusters and are located at
the density peak of member galaxy in space with a small difference
between their photometric redshift and spectroscopic redshift
($\Delta z <$ 0.02).

Note that the SDSS fiber spectrograph only samples the light
within the central 3" of galaxies (Strauss et al. 2002). A large aperture bias for
selection of nearby E+A galaxies can be avoided by a low redshift
cut of $z>0.05$ (Goto 2007). The star formation activity in BCGs tends to
happen in the central region (e.g., Bildfell et al. 2008; Pipino et al. 2009). 
Probably it holds true for the post-starburst E+A galaxies as well. Our E+A galaxies as BCGs
have a weaker \Hd absorption line. Such galaxies could be roughly
classified as weak E+A galaxies by Poggianti et al. (2004), which
historically had much less violent star formation than general E+A
galaxies in the field.

E+A (post-starburst) galaxies are really very rare among BCGs.
Only five of 8,812 BCGs show the significant E+A features. This is
expected, since the BCGs are the dynamically oldest objects of the
clusters, whereas the E+A galaxies experienced starburst within
about 1~Gyr.

We compared the physical properties of these E+A BCGs with those
of two control samples. One is the catalog of 564 general E+A
galaxies found mostly in the field by Goto (2007), and the other
includes 154 quiescent BCGs with definitely no star formation
(i.e. [O II] EW$>$0\AA~and \Ha~EW$>$0\AA) but with a similar
stellar mass of these E+A BCGs ($11.25<\log M_{\ast}<11.75$) in
the redshift of 0.25$<$z$<$0.37. We collected the 4000\AA~break
strength, $D_n(4000)$, the Balmer line absorption index (Worthey et al. 1997),
$H{\delta}_A$, and the total stellar mass (log$M_{\ast}$) of the
objects in control samples from the MPA/JHU catalogs. Not all
objects have these parameters estimated already there: 543 out of
564 general E+A galaxies of Goto (2007) have all these parameters on
line, and four of five E+A BCGs have total stellar mass estimated.
We compared the $D_n(4000)$, stellar mass and the line absorption
index, $H{\delta}_A$, of E+A BCGs with those of two control
samples (Figure 2).
E+A BCGs have distinct different locations in those parameter space
from the field E+A galaxies. They are usually more massive, and have
larger $D_n(4000)$ and smaller absorption line index $H{\delta}_A$,
which indicates their different star formation history. The star
formation in the field E+A galaxies may be triggered by interaction of
gas-rich galaxies (e.g., Goto 2005), while star formation in BCGs is related to
cluster cooling flows (e.g., Fabian 1994).

E+A BCGs have almost the same $D_n(4000)$ distribution as
quiescent BCGs. However, E+A BCGs have slightly larger
$H{\delta}_A$ than quiescent BCGs with similar stellar masses,
which indicates that the mean stellar ages in E+A BCGs may be less
than those in normal BCGs. They obviously experienced recent star
formation. E+A galaxies represent an important stage of overall
galaxy evolution. Detailed studies of these E+A BCGs and their
environment may help to understand why some BCGs had the starburst
and why the starburst was quenched later.

\acknowledgements

We thank Prof. Shude Mao for useful
discussions and the anonymous referee for valuable suggestions and
comments. The authors are supported by the Liaoning Educational
Foundation of China (No.  2009A646) and the National Natural
Science Foundation (NNSF) of China (10773016, 10821061, and
1083303) and the National Key Basic Research Science Foundation of
China (2007CB815403).
Funding for the creation and distribution of the SDSS Archive has
been provided by the Alfred P. Sloan Foundation.

\clearpage
\begin{table*}
\scriptsize
\begin{minipage}{178mm}
\setlength{\tabcolsep}{0.038in}
\caption[]{Basic parameters for five E+A BCGs}
\begin{center}
\begin{tabular}{lcrrccccccccc}
\hline
\mc{1}{c}{Cluster Name} &
\mc{1}{c}{BCG: R.A.} &
\mc{1}{c}{BCG: Dec.} &
\mc{1}{c}{BCG: z} &
\mc{1}{c}{$\rm S/N$ } &
\mc{1}{r}{$\rm {EW{\rm (H_{\delta})}}$ } &
\mc{1}{r}{$\rm {EW{([\rm O II]) }} $ } &
\mc{1}{r}{$\rm {EW{\rm (H_{\alpha})}}$ } &
\mc{1}{c}{log $\rm M_{\ast}$} &
\mc{1}{c}{$D_n$(4000) } &
\mc{1}{c}{$H{\delta}_A$ } \\

\mc{1}{c}{(1)} & \mc{1}{c}{(2)} & \mc{1}{c}{(3)} & \mc{1}{c}{(4)} & \mc{1}{c}{(5)} & \mc{1}{c}{(6)} & \mc{1}{c}{(7)} & \mc{1}{c}{(8)} & \mc{1}{c}{(9)}  & \mc{1}{c}{(10)} & \mc{1}{c}{(11)}
\\

\hline
WHLJ110352.5$+$042234  & 165.96890  & 4.37639   & 0.3218  & 6.51& 3.92$\pm$0.84 & 4.39$\pm$2.04   & -2.97$\pm$0.64 & 11.52 &1.93$\pm$0.09  & 2.19$\pm$1.99\\
WHLJ122626.3$+$135107  & 186.60950  & 13.85196  & 0.2526  & 5.19& 3.30$\pm$1.09 & 0.45$\pm$2.22   & -0.56$\pm$0.58 & 11.48 &1.82$\pm$0.10  & 1.10$\pm$2.68\\
WHLJ133720.1$+$160830  & 204.33380  & 16.14181  & 0.3790  & 5.47& 2.57$\pm$0.76 & -1.26$\pm$1.33  & -0.39$\pm$0.83 & ... &1.98$\pm$0.07  & 1.00$\pm$1.86\\
WHLJ144313.4$+$280032  & 220.80580  & 28.00913  & 0.3097  & 5.38& 2.95$\pm$0.87 & -0.92$\pm$1.71  & -0.87$\pm$0.73 & 11.58 &1.99$\pm$0.08  & 1.79$\pm$2.14\\
WHLJ145942.3$+$192405  & 224.88609  & 19.37866  & 0.2772  & 5.30& 2.82$\pm$0.91 & -2.12$\pm$2.01  & -1.00$\pm$0.73 & 11.52 &1.75$\pm$0.08  & 3.02$\pm$2.28\\

\hline
\end{tabular}
\end{center}
{Note:
Col:(1) SDSS-WHL Cluster Name.
Col:(2) BCG R.A.(J2000.0) in unit of degree.
Col:(3) BCG Dec.(J2000.0) in unit of degree.
Col:(4) The spectroscopic redshift of the BCG.
Col:(5) The signal-to-noise (S/N) of the whole spectrum.
Col:(6) The equivalent width of ${\rm H_{\delta}}$ in unit of ${\rm \AA}$.
Col:(7) The equivalent width of ${\rm [O II]}$$\lambda$3727, in unit of ${\rm \AA}$.
Col:(8) The equivalent width of ${\rm H_{\alpha}}$ in unit of ${\rm \AA}$.
Col:(9) Logarithm of total stellar mass, in unit of $M_{\odot}$.
Col:(10) The amplitude of the 4000 Balmer break, $D_n$(4000).
Col:(11) The absorption line index $H{\delta}_A$.
}
\label{tab:bcg}
\end{minipage}
\end{table*}

\clearpage
\begin{figure}
\epsscale{1.0}
\figurenum{1}
\plotone{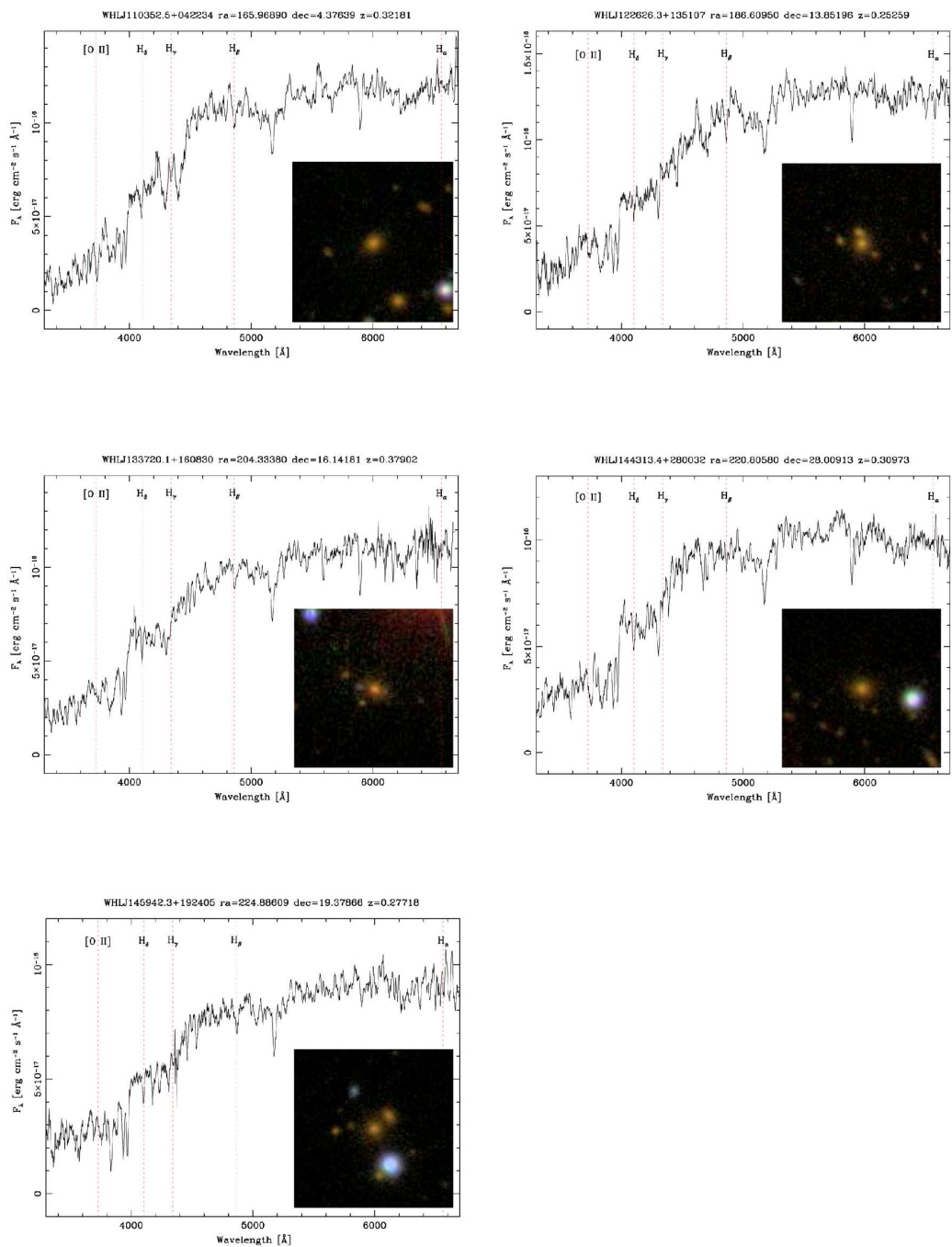}
\caption{
Spectra and color images of five
  E+A BCGs. Each spectrum is shifted to the restframe wavelength,
  corrected for the Galactic extinction, and smoothed using a 15
  \AA~box. The size of color images corresponds
  200~kpc$\times$200~kpc.
}
\label{f1.eps}
\end{figure}

\clearpage
\begin{figure}
\epsscale{.99}
\figurenum{2}
\plotone{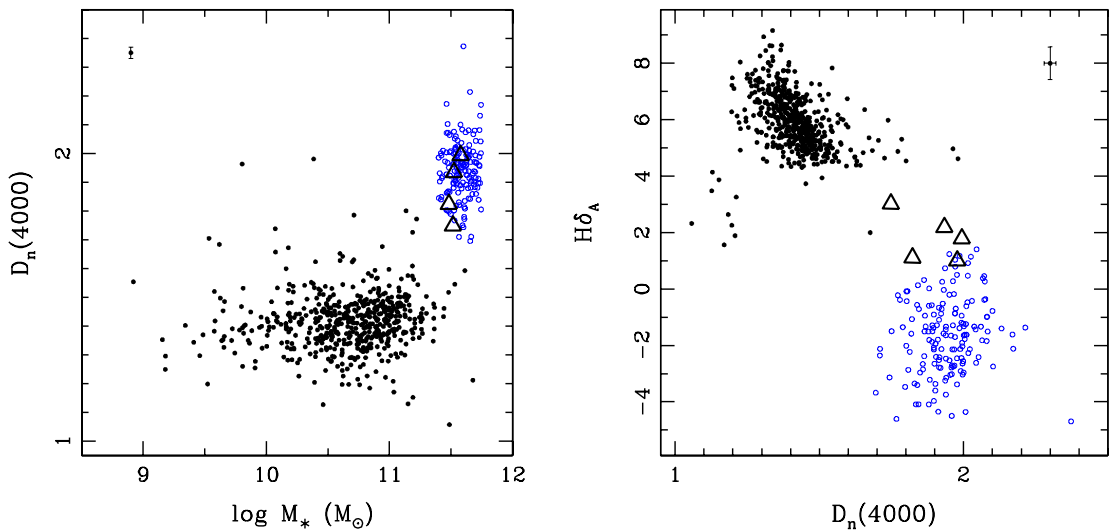}
\caption{
E+A BCGs (triangles) are compared
  to the 543 field E+A galaxies (dots, from Goto 2007) and 154 normal
  quiescent BCGs with similar masses (circles). Apparently, E+A BCGs
  have not only different mass from the field E+A galaxies ({\it left
    panel}) but also different occupations in the parameter space of
  $D_n(4000)$ and $H{\delta}_A$ ({\it right panel}) from the field E+A
  galaxies and quiescent BCGs.
}
\label{f2.eps}
\end{figure}

\end{document}